\documentclass[12pt,preprint]{aastex}

\received{28 May 2005}
\def\Pdot{$\dot P$}
\def\Msun{$M_\odot$}
\slugcomment{}
\shortauthors{Kepler et al.}
\shorttitle{Evolutionary timescale of G~117-B15A}
\begin{document}

\title{Measuring the evolution of the most stable optical clock G~117-B15A}
\author{S.O. Kepler, J.E.S. Costa, B.G. Castanheira}
\affil{Instituto de F\'{\i}sica da UFRGS, 91501-900 Porto Alegre, RS - Brazil,
kepler@if.ufrgs.br}
\author{
D.E. Winget,
Fergal Mullally,
R.E. Nather, 
Mukremin Kilic \&  Ted von Hippel}
\affil{Department of Astronomy and McDonald Observatory, 
University of Texas, Austin, TX 78712 - U.S.A.}
\author{Anjum S. Mukadam}
\affil{Department of Astronomy, University of Washington, Seattle, WA\,-\,98195-1580, U.S.A.}
\author{Denis J. Sullivan}
\affil{School of Chemical And Physical Sciences,
                      Victoria University of Wellington, New Zealand}

\begin{abstract}
We report our measurement of the
rate of change of period with time ($\dot{P}$) for the 215\,s periodicity in the pulsating white dwarf G~117--B15A,
the most stable optical clock known.
After 31 years of observations, we have finally obtained a 4\,$\sigma$ measurement
$\dot P_{\mathrm{observed}} = (4.27 \pm 0.80) \times 10^{-15} \,\rm s/s$. 
Taking into account the proper-motion effect of
$\dot P_{\mathrm{proper}} = (7.0 \pm 2.0) \times 10^{-16} \,\rm s/s$,
we obtain a rate of change of period with time of
$\dot P = (3.57 \pm 0.82) \times 10^{-15} \,\rm s/s$.
This value is consistent with the cooling rate in our white dwarf models
only for cores of C or C/O. With the refinement of the models,
the observed rate of period change can be used to accurately
measure the ratio of C/O in the core of the white dwarf.
\end{abstract}


\section{Introduction\label{intro}}
\object[G 117-B15A]{G~117--B15A} is a 
pulsating white dwarf with a hydrogen atmosphere,
a DAV, also called a ZZ Ceti star
\citep{M79}.
These stars
show multi-periodic non-radial $g$-mode pulsations 
that can be used to measure
their internal properties and rate of evolution.

\citet{MR76}
found the star was variable,
and \citet{K82}
studied its light curve, finding 6 simultaneous pulsations. 
The dominant mode has a
period of
215~s, a fractional amplitude of 22 mma (milli-modulation amplitude = 1/1.086 milli magnitude),
and is stable in amplitude
and phase.
The other smaller pulsation modes vary in amplitude
from night to night \citep{K95},
suggesting 
the presence of unresolved components.
Because the DAVs appear to be normal stars except for their
variability \citep{Robinson79, Bergeron95, Bergeron04},
i.e., an evolutionary stage in the cooling of all white dwarfs,
it is likely that the DAV structural properties are representative
of {\it all}  DA white dwarfs.

\citet{Mukadam04a} and \cite{Fergal} discovered 46 new ZZ Cetis 
and more than doubled the number
of known variables, using $T_{\mathrm{eff}}$ and $\log$~g values 
derived from the optical spectra obtained by
the Sloan Digital Sky Survey \citep{Scot}.
While \citet{Bergeron95, Bergeron04} find the ZZ Ceti instability 
strip to be pure,
i.e., contain no non-variable stars, 
\citet{Mukadam04a,Mukadam04b,Fergal} find
several 
stars inside the instability strip for which they could detect
no variability above their detection threshold.
It is necessary to obtain $S/N\geq 50$ spectra of the non-variables 
within the strip
and re-analyze their $T_{\mathrm{eff}}$ and $\log$~g values,
and also to  obtain
additional time series photometry on these stars, to ensure that they
are not low amplitude variables, and re-study the purity of the 
instability strip.

We report our continuing study
of the star G~117--B15A, also called RY LMi and WD~0921+354,
one of the hottest of the ZZ Ceti stars.
We expect that the rate of change of a pulsation period with time for $g$-mode pulsations
in white dwarf stars
to be directly related to its evolutionary timescale
\citep{WHVH},
allowing us to infer the age of a cool white dwarf since its
formation.
We have been observing the star since 1974 to measure the
rate of period change with time ($\dot P$)
for the largest amplitude periodicity.
Using all the data obtained from 1974 through 1999, \citet{K00} 
arrived at a determination of
$\dot P= (2.3 \pm 1.4) \times 10^{-15}\,{\rm s/s}$.

G~117--B15A was the first pulsating white dwarf to have its main
pulsation mode index identified. The 215~s mode is an $\ell=1$, 
as determined by comparing the
ultraviolet pulsation amplitude (measured with the Hubble
Space Telescope) to the optical amplitude \citep{R95}.
\citet{Kotak} confirm the $\ell$
measurement for the P=215~s pulsation, and show that the
other large amplitude modes, at 271~s and 304~s,
show chromatic amplitude changes that do not fit the
theoretical models, using time-resolved spectra obtained
at the Keck Telescope.
\citet{R95}, and \citet{KAV}
derive $T_{\mathrm{eff}}$ near 12,400~K, 
while \citet{Bergeron95,Bergeron04}
using a less efficient model for convection, 
derive $T_{\mathrm{eff}}$=11,630~K.

\citet{K84} demonstrated that the observed
variations in the light curve of G~117--B15A 
are due to non-radial {\it g}-mode pulsations.
\citet{K00} show the models
predict the effect of radius change due to still ongoing
contraction to be an order of magnitude smaller than the
cooling effect on the rate of period change.

G~117--B15A is proving to be a useful laboratory for particle
physics \citep{Isern}.
\citet{Corsico2001}
calculated the limit on the axion mass compatible with the
then observed upper limit to the cooling, 
showing $m_a \cos\beta\leq 4.4~\mbox{meV}$ and \citet{K04}
demonstrates axion cooling would be dominant over neutrino
cooling for the lukewarm white dwarf stars for axion masses
of this order.
\citet{Biesiada}
show that the $2\sigma$ upper limit published in \citet{K00}
limits the string mass scale $M_S \geq 14.3~\mbox{TeV}/c^2$ for 6 dimensions,
from the observed cooling rate and the emission of Kaluza-Klein gravitons,
but
the limit is negligible for higher dimensions.
\citet{Benvenuto04} show the
observed rates of period change can also be used to constrain the
dynamical rate of change of the constant of gravity $\dot G$.

\citet{Bradley96,Bradley98} used the mode identification and the 
observed periods of the 3 largest known
pulsation modes to derive a hydrogen layer mass lower limit
of $10^{-6}\,M_*$, and a best estimate of $1.5 \times 10^{-4}\,M_*$,
assuming $k=2$ for the 215~s mode, and 20:80 C/O core mass. The
core composition is constrained mainly by
the presence of the 304~s pulsation.
\citet{Benvenuto02} show the seismological models with time-dependent
element diffusion are only consistent
with the spectroscopic data if the modes are the $\ell=1$, k=2, 3, and
4,
and deduces $M=0.525~M_\odot$, $\log(M_H/M_*)\geq -3.83$ and
$T_{\mathrm{eff}}=11\,800$~K. 
%similar to those by \citet{KA00}.
Their best model predicts: parallax $\Pi$=15.89~mas,
$\dot{P}=4.43 \times 10^{-15}$~s/s, for
the P=215~s,
$\dot{P}=3.22 \times 10^{-15}$~s/s, for
the P=271~s, and
$\dot{P}=5.76 \times 10^{-15}$~s/s, for
the P=304~s periodicities.
%SSS we should quote these numbers in section 5.

\section{Observations}

\cite{K00} reported on the observations from 1974 to 2000.
We report in this paper
additional 19.3~h of time series photometry in 2001,
30.6~h in 2002, 24.4~h in 2003, 4~h in 2004, and 13.6~h in 2005,
most using
the Argos prime-focus CCD camera 
\citep{NA}
on the $2.1$~m
Otto Struve telescope at McDonald Observatory.

We observed the light through a BG40 filter to maximize the
signal-to-noise ratio (S/N) because the pulsation amplitudes are 
small (2\%),
the star is faint
(V=15.52, \citealt{EG}), and also because the non-radial
{\it g}-mode light variations have the same phase in all colors
\citep{RKN} but
the amplitudes decrease
with wavelength. For example, a filter-less observation with
Argos gives an amplitude around 40\% smaller for G~117--B15A.

\section{Data Reduction}
We reduce and analyze the data in the manner described by 
\citet{N90},
and \citet{K93}.
We bring all the data to the same fractional amplitude scale,
and the times 
from UTC
to the uniform Barycentric Julian Coordinated Date
(TCB) scale, using JPL DE405 ephemeris \citep{Standish98,Standish04}
to model Earth's motion.
We compute Fourier transforms for each individual run,
and verify that the  main pulsation at 215~s
dominates each data set and has an amplitude stable up to 15\%,
our uncertainty in amplitude due to the lack of accurate
time- and color-dependent extinction determination.

\section{Time Scale for Period Change}

As the dominant pulsation mode at P=215~s has a stable frequency and amplitude
since our first observations in 1974, we can calculate the
time of maximum for each new run and look for deviations.

We fit our observed time of maximum light to the equation:
\[(O-C) = \Delta E_0 + \Delta P \cdot E
+ \frac{1}{2} P \cdot \dot P \cdot E^2\]
where $\Delta E_0 = (T_{max}^0 - T_{max}^1)$, $\Delta P = (P - P_{t=T_{max}^0})$, $E$ is the epoch of the time of maximum, i.e, the integer number of cycles
after our first observation,
$T_{max}^0$ is the time of maximum assumed,
$T_{max}^1$ is the time of maximum that best fit the parabola,
$P$ is the period that best fit the parabola, and
$P_{t=T_{max}^0}$ is the period assumed at $T_{max}^0$
{\footnote{Fitting the
whole light curve with a term proportional to
$\sin\left[\frac{2\pi}{\left(P+\frac{1}{2}\dot{P}\right)}t + \phi\right]$ by non-linear least squares gives unreliable uncertainty estimates
and the alias space in P and $\dot{P}$ is extremely dense
due to 31\,yr data set span \citep{Darragh,Costa99}.
Our non-linear least squares result is $\dot{P}=(3.38\pm 0.0013) \times 10^{-15}$s/s.}}.
The times of maxima are calculated by a linear least-squares
fit of the light curve of each night 
to a sum of the six detected frequencies. They are shown in 
Table~\ref{tmax}.

\begin{figure}
\resizebox{\hsize}{!}{\includegraphics*[angle=0]{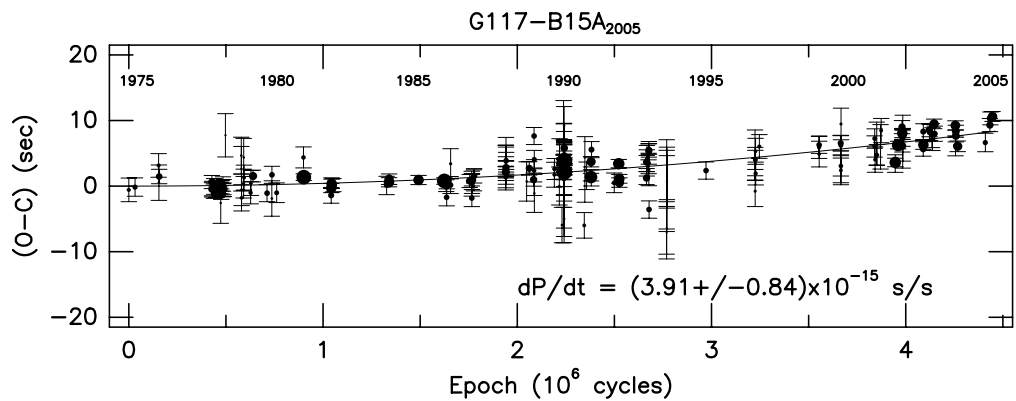}}
\caption{
{\bf (O-C)}: ({\bf O}bserved minus {\bf C}alculated times of maxima) for
the 215~s pulsation of G~117-B15A. The size of each point is proportional
to its weight, i.e., inversely proportional to the uncertainty
in the time of maxima squared. We show 2$\sigma$ error bars for each point,
and the line shows our best fit parabola to the data.
The error bars plotted are those before adding the external
uncertainty of 1s quadratically,
discussed in the text.
The fact the line does not overlap these error bars are
a demonstration they are underestimate.
Note that as the period of pulsation is 215.197s, the whole plot shows
only $\pm 36\deg$ in phase.}
\label{Figure1}
\end{figure}

In Figure~\ref{Figure1}, we show the O--C timings 
after subtracting the correction to period and epoch,
and our best fit curve through the data.
The size of each point is proportional to its weight,
i.e., inversely proportional to the square of
uncertainty in phase. The error bars plotted are $\pm 1\sigma$.
From our data through 2005, we obtain a new value for the epoch of maximum,
$T_{max}^0 = 244\,2397.9175141 \,{\rm TCB} \pm 0.41 \,\rm s$,
a new value for the period, $P = 215.197 388 8 \pm 0.000 000 4 \,\rm s$,
and most importantly, a rate of period change of:
\[\dot P = (4.27 \pm 0.80) \times 10^{-15} \,\rm s/s.\]

\begin{figure}
\resizebox{\hsize}{!}{\includegraphics*[angle=0]{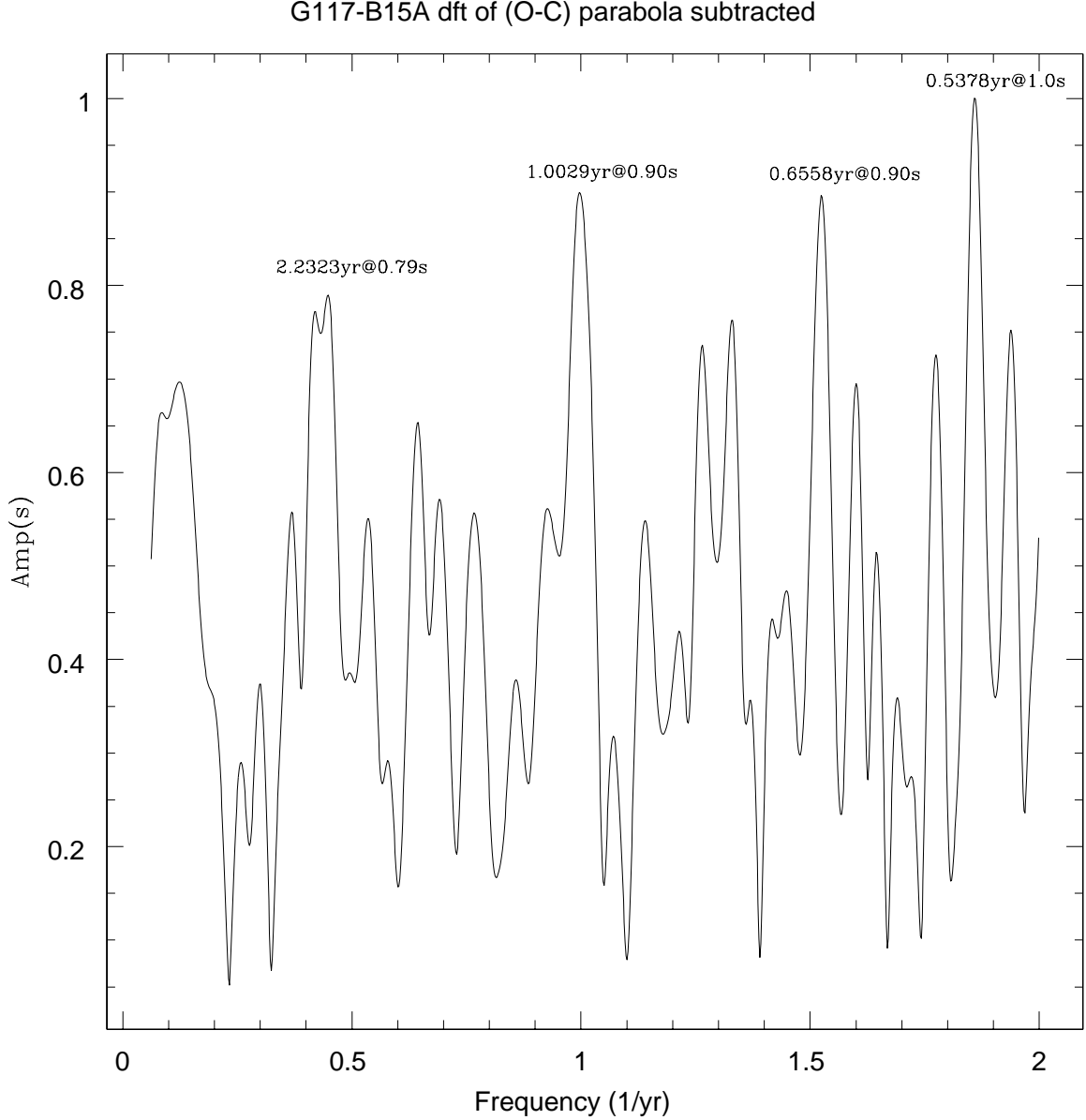}}
\caption{
Fourier transform of the (O-C): ({\bf O}bserved minus {\bf C}alculated times of maxima) for
the 215~s pulsation of G~117-B15A, after we subtract the parabola.
Kepler's law (1616) shows that 
$a_*\simeq \frac{M_P}{M_*^{2/3}}\left(\frac{G}{4\pi^2}\right)^{1/3}
P^{2/3}$, considering $M_P \ll M_*$, from which
we
get 
$a_* \mbox{(light-second)} \simeq 0.565 \frac{M_P}{M_J} P^{2/3}\mbox{(yr)}$
i.e., a 1s light travel time, as the limit shows, for planets
with masses between 1.6~$M_J$ and 0.27~$M_J$=87~$M_\oplus$, for periods between
6~months and 7~yr, respectively. The highest peak, at a frequency of
1.85~$\mbox{yr}^{-1}$, corresponding to a period of 6.5~months,
is extremely hard to study in our data set, as we normally
observe for only 3 months separated by 1 year. These  variations
can be caused by beating of very closely spaced pulsations
(\cite{K95}), and we must take into account that
multiplets cause non-sinusoidal variations, as
the triplet in G~226-29 studied by \cite{K83}.
}
\label{Figure2}
\end{figure}

We use linear least squares to make our fit, with each point weighted inversely
proportional to the uncertainty in the time of maxima for each 
individual run squared.
We quadratically add an additional 1s of uncertainty to the time of maxima
for
each night to account for external uncertainty caused perhaps
by the beating of possible small amplitude pulsations \citep{K95}
or the small modulation seen in Figure~\ref{Figure1}.
The amplitude, 1s, is chosen from the Fourier transform of the
(O-C) shown in Figure~\ref{Figure2}, and is in agreement
with \citet{KP} conclusion that the Fourier analysis of the
Times of Arrival (TOA) gives unbiased information
about the noise.
Such external uncertainty is also consistent with \citet{Splaver} who
show that the true uncertainties of the
times of arrival of the milli-second pulsars are generally larger
than the formal uncertainties, and that a quadratic term is
added to them to fit the observations.

\section{Discussion}
  While it is true that the period change timescale can be proportional to
   the cooling timescale, it is also possible that other phenomena with
   shorter timescales can affect $\dot P$. The cooling timescale is the longest
   possible one.

As a corollary, if the observed $\dot P$ is low enough
   to be consistent with evolution, then other processes, such as perhaps a
   magnetic field or
diffusion induced changes in the boundary layers,
are not present at a level sufficient to affect $\dot P$.

For the first time we also report on the search for the rates
of period changes for the other relatively large amplitude
modes of G~117--B15A,
$dP/dt = (36.0 \pm 7.2) \times 10^{-15}$~s/s for the 270s periodicity, and
$dP/dt=(74 \pm 15) \times 10^{-15}$~s/s for the 304s periodicity.
They are much larger than the one derived for the main
pulsation, which has a region of period formation deeper in the
core, but are in line with the measurements for the
274s modes in ZZ Ceti \citep{r548}. The models to date,
without rotation, differential rotation, and magnetic fields, do not
explain such values,
or the chromatic amplitude changes
reported by \citet{Kotak}. \citet{K95} show the 270s and 304s
periodicities have significant amplitude changes, indicative of
multiple components or resonances.

% QQQ I think we are sitting on some wonderful measurements and we need an elaborate discussion on this.
% To me, the most important point of the paper is that you have 3 different Pdot measurements for
% for 3 different modes in the same star. So revealing.
% The one thing that scares me is that we could simply be measuring the timescales at which
% these modes interact with one another. We assume otherwise, and we could be wrong.
% We need a lot of prudence in presenting our results.

\subsection{Theoretical Estimates and Corrections}

\subsubsection{Proper Motion}
\begin{figure}
\resizebox{\hsize}{!}{\includegraphics*[angle=0]{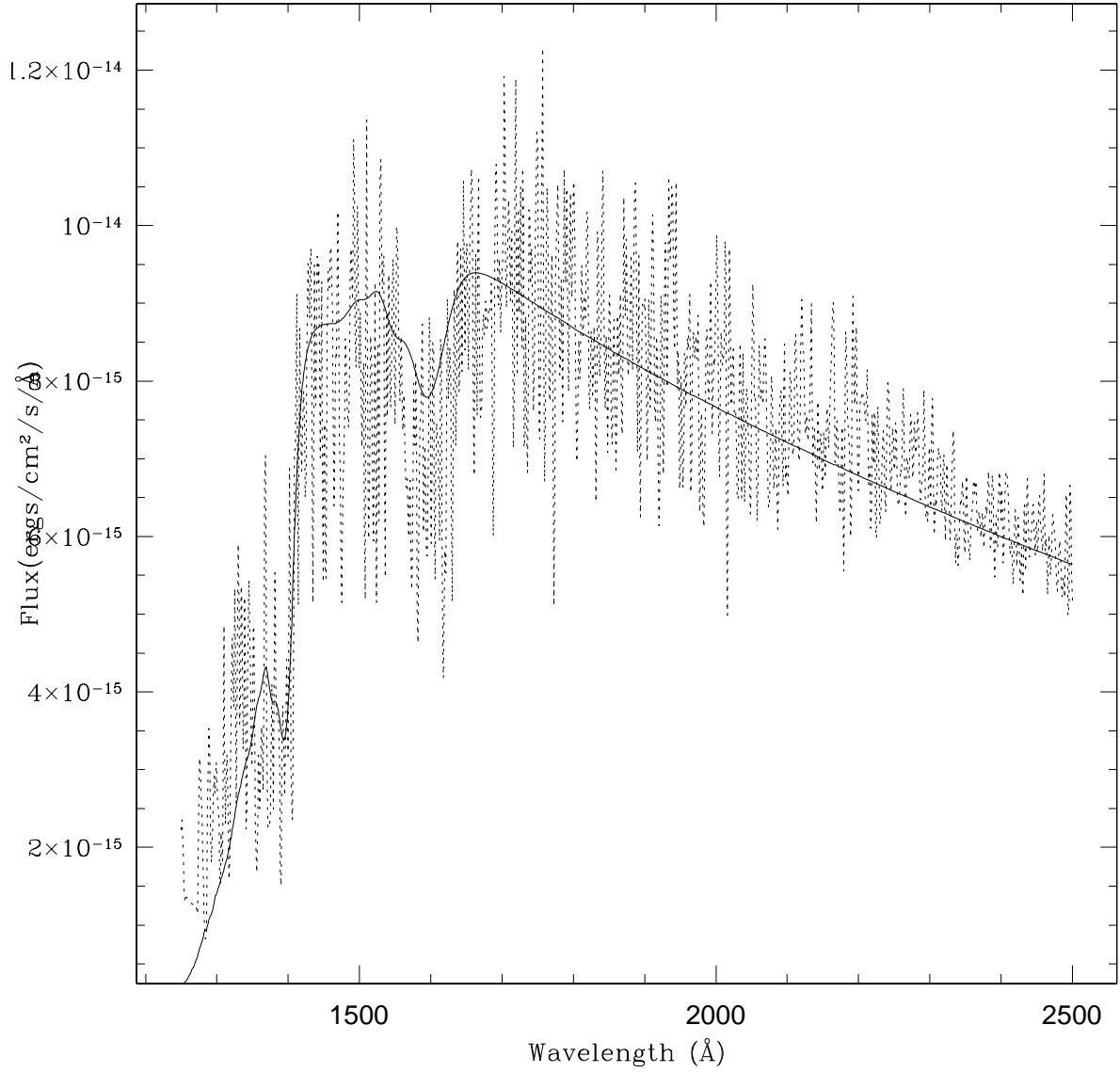}}
\caption{Fit of a model atmosphere with $T_{\mathrm{eff}}=12\,000$\,K,
$\log g=8.0$, d=67\,pc, to the HST FOS spectra of G~117-B15A.
}
\label{hst}
\end{figure}

\citet{Pajdosz95} discusses the influence of the proper motion of the star
on the measured $\dot P$:
\[\dot P_{\mathrm{obs}} = \dot P_{\mathrm{evol}}\left(1+v_r/c\right)
+ P\dot v_r/c\]
where $v_r$ is the radial velocity of the star. Assuming
$v_r/c \ll 1$
he derived
\[\dot P_{\mathrm{pm}} = 2.430 \times 10^{-18} P[s]
\left(\mu[\,"/yr]\right)^2 d[{\mathrm{pc}}]\]
where $\dot P_{\mathrm{pm}}$ is the effect of the proper motion
on the rate of period change, $P$ is the pulsation period,
$\mu$ is the proper motion and $d$ is the distance.
The proper motion, $\mu=0.136 \pm 0.002\,"/{\mathrm{yr}}$,
and the parallax,
$\Pi=(0.0105 \pm 0.004)\,"$,
are given by \citet{vanAltena95}.
But the parallax has a large uncertainty and does not agree with
other estimates of the distance.
\citet{Munn} measured the same proper motion for G~117-B15A and
G~117-B15B:
$\mu_\alpha=(-146 \pm 2.6)$~mas/yr, and
$\mu_\delta=(-1.0 \pm 2.6)$~mas/yr, 
including 5 USNO-B epochs and SDSS positions.
Considering
$V^A=15.50\pm 0.02$ \citep{Silvestri},
%V=15.45 \citet{Silvestri}, 15.52 \citet{EG}, 15.5 \citet{HD80}
%v_r(B)=2.2\pm 9.4 km/s
if we use \citet{Bergeron04} estimate of
$M_V^A=11.70\pm 0.06$\footnote{We estimate the uncertainty in
the absolute magnitude from the 300~K external
uncertainty in Bergeron's temperatures.} derived from their optical spectra,
we obtain a distance of $58\pm 2$~pc, equivalent to a spectroscopic
parallax of $0.0172\pm 0.0005$".
But G~117-B15A has International Ultraviolet Explorer (IUE) 
and Hubble Space Telescope (HST) Faint Object
Spectrograph (FOS) flux calibrated spectra
\citep{KAV}
that can
also be used to estimate the distance, if we use the evolutionary
models of \citet{Wood} or \citet{laplata} to estimate the
radius from $T_{\mathrm{eff}}$ and $\log g$ and fit
the observed fluxes to a model atmosphere calculated
by Detlev Koester, similar to that described in
\citet{Finley}, but with an ML2/$\alpha$=0.6 convection
description consistent with \citet{Bergeron95}
and \citet{KH}.
The IUE spectra
re-calibrated according to the New Spectroscopic Image
Procession System (NEWSIPS) data reduction by NASA
was published by \citet{iue}.
From the IUE spectra we get a distance of $d=59\pm 6$~pc.
The HST spectra, shown in Figure~\ref{hst},
fits a distance of $d=67\pm 5$~pc. Table~\ref{dist}
lists all the available distances to G~117-B15A.
\begin{table}[ht!]
\begin{center}
\caption{G~117-B15A distance determinations}
\label{dist}
\begin{tabular}{lr}
{\bf Method}&{\bf distance (pc)}\\
Parallax&$95\pm 37$\\
Spectroscopic parallax&$58\pm 2$\\
IUE flux&$59\pm 5$\\
HST flux&$67\pm 4$\\
Seismology from \citet{Bradley98}&61\\
Seismology from \citet{Benvenuto02}&63\\
{\bf Mean}&{\bf $67\pm 14$}
\end{tabular}
\end{center}
\end{table}

Taking the average value of proper motion and distance, we estimate
\[\dot P_{\mathrm{pm}}= (7.0 \pm 2.0) \times 10^{-16}\,\rm s/s\]
and
\[\dot P = 
\dot P_{\mathrm{observed}} -
\dot P_{\mathrm{pm}} = 
(3.57 \pm 0.82) \times 10^{-15} \,\rm s/s\]

\subsection{Pulsation Models}
We compare the measured value
of \Pdot\ with
the range of theoretical values derived from realistic evolutionary models
with $C/O$ cores
subject to {\it g}--mode pulsations
in the temperature range
of G~117--B15A.
The adiabatic pulsation calculations
of 
\citet{Bradley96}, and
\citet{Brassard92,Brassard93}, which allow for mode trapping, give
\Pdot $\simeq 2-7 \times 10^{-15} \,\rm s/s$
for the $\ell=1$, low {\it k} oscillation observed.
\citet{Benvenuto04} estimated the theoretical $\dot P$ for
the three modes of G~117-B15A, even allowing for a dynamical
change on the gravity constant, as we show in section~\ref{intro}.
The observed $P/\dot P =1.9 \times 10^9$ yr,
equivalent to 1~s change in period
in 8.9 million years, is within the theoretical
predictions and very close to it. We have therefore
measured a rate consistent with the evolutionary time scale
for this lukewarm white dwarf.

\subsubsection{Core Composition}
For a given mass and internal temperature distribution,
theoretical models show that the rate of period change
increases if the mean atomic weight of the core is increased,
for models which have not yet crystallized in their interiors.
As the evolutionary model cools, 
its nucleus crystallizes due to Coulomb
interactions between the ions
\citep{Lamb}, and crystallization  slows down
the cooling by the release of latent heat.
\citet{Mike99} describe the effect of 
crystallization on the pulsations of white dwarf stars,
but G~117--B15A is not cool enough to have a crystallized core
\citep{W97}, or even for the convective coupling
described by \citet{Fontaine01} to occur.

The heavier the particles
that compose the  nucleus of the white dwarf, the faster it
cools.
The best estimate of mean atomic weight $A$ of the core comes from the
comparison of the observed $\dot P$ with values from an
evolutionary sequence of white dwarf models. 
\citet{Brassard92} computed the rates of period changes for
800 evolutionary models with various masses,
all with carbon cores but differing He/H surface layer masses, 
obtaining values similar to those
of \citet{W81}, 
\citet{Wood88},
and \citet{Bradley91}.
The average value of $\dot P$ for all
$\ell=1$, 2 and 3 modes with periods around 215~s in models with an
effective temperature around 13,000~K, and a mass of 0.5~\Msun, is:
$\dot P(\mbox{C core}) =  (4.3 \pm 0.5) \times 10^{-15}\,\rm s/s.$
\citet{Benvenuto04} C/O models give 
$\dot P(\mbox{C/O core}) =  (3-4) \times 10^{-15}\,\rm s/s.$
Using a Mestel-like cooling law \citep{Mestel52, Kawaler86}, i.e.,
$\dot T \propto A$, where $A$ is the mean atomic weight in the core,
we can write:
\[\dot P(A) =  (3-4) \times 10^{-15}\,\frac{A}{14} \,\rm s/s.\]
The observed rate of period change is
therefore consistent with a C or C/O core. The largest uncertainty comes
from the models.

\subsubsection{Reflex Motion}
The presence of an orbital companion
could contribute to the period change we have detected.
When a star has an orbital companion, the variation of its line-of-sight
position with time produces a variation in the time of arrival of
the pulsation maxima, by changing the light travel time between the star
and the observer by reflex motion of
the white dwarf around the barycenter of the system.
\citet{K91} estimated a contribution to $\dot{P}$ caused by
reflex orbital motion of 
the observed proper motion companion of G~117--B15A in
their equation (10) as:
\[\dot{P}_{\mathrm{orbital}} = \frac{P_{\mathrm{pul}}}{c} 
\frac{GM_B}{a_T^2}
= 1.97 \times 10^{-11} P_{\mathrm{pul}}\,
\frac{M_B/M_\odot}{(a_T/AU)^2}\,{\mathrm{s/s}}\]
where
$a_T$ is 
the total separation and $M_B$ is the mass of the companion star.
In the above derivation
they have also assumed
the orbit to be nearly edge on to give the largest effect possible.
Only the acceleration term parallel to the line of sight
contributes to $\dot{P}$.
Even though G~117-B15A and 
\object[G 117-B15B]{G~117--B15B} are a common proper motion pair
\citep{Giclas, vanAltena}, there is no other
evidence they form a real binary system. 
%\citet{Silvestri}
%measured the radial velocity of G~117-B15B, but assumes it forms
%a wide binary system with G~117-B15A.
\citet{Kotak} classifies G~117-B15B as an M3Ve from its spectra,
obtained with the 10~m Keck I telescope,
and measured $\log g\simeq 4.5$ and $T_{\mathrm{eff}}\simeq 3400$~K.
With 
$V^B=16.1$  \citep{HD80,vanAltena}
and $M_V^B\simeq 10.4\pm 0.9$
\citep{Lang}
with the uncertainty coming from a possible misclassification
in the spectral type of G~117--B15B
from M3V to M4V. Such uncertainty arises
from the TiO and CaOH bands
seen by \citet{Kotak} and the (B-V)=1.63 \citep{HD80}.
Its spectroscopic distance is $d^B=138\pm 47$~pc. 
G~117-B15B
is chromospherically active \citep{Kotak} and a flare star
\citep{NA}. 
G~117-B15A and B were imaged by
SDSS DR3, with
$u_A=15.92,\ g_A=15.54,\ r_A=15.64,\ i_A=15.80,\ z_A=16.06$, and
$u_B=19.38,\ g_B=16.95,\ r_B=15.46,\ i_B=13.97,\ z_B=13.16$, which
also measure a separation of 13.4".
They were also observed by 2MASS \citep{Cutri},
with $J_A=15.599 \pm 0.065,\ H_A=15.614 \pm 0.12$, and
$J_B=11.599 \pm 0.018,\ H_B=11.156 \pm 0.12$, and
separation of 13.4".
Using \citet{Pickles} main sequence spectral templates convolved
with SDSS filters, our best fitting template is an M4V,
with $M_V^B=11.54$, and a distance of 82~pc.
Using \citet{Hawley} calibration for the infrared colors,
G~117-B15B is consistent with the M3V spectral type, 
and the (i-z) colors correspond to $M_i^B=10.33$,
and a distance of 54 parsecs.
Using the (i-J) colors, $M_J^B=10.05$, we obtain a distance of 61 parsecs.
The parallaxes $\pi^A=11 \pm 5$~mas,
$\pi^B=5 \pm 7$~mas
\citep{vanAltena},
and $\pi^B=4 \pm 7$~mas \citep{HD80}
are too uncertain to be used to study the binary nature.
Table~\ref{dB}
lists all the available distances to G~117-B15B.
\begin{table}[ht!]
\begin{center}
\begin{tabular}{lr}
{\bf Method}&{\bf distance (pc)}\\
Parallax&$200\pm 280$\\
Spectroscopic parallax&$138\pm 47$\\
SDSS colors&$82$\\
(i-z) colors&$54$\\
(i-J) colors&$61$\\
{\bf Mean}&{\bf $107 \pm 62$}
\end{tabular}
\end{center}
\caption{Distance determinations to G~177-B15B}
\label{dB}
\end{table}
The mass of an M3.5V should be around 
$(0.30\pm 0.03)~M_\odot$ \citep{Lang}.
With a separation around 13.4 arcsec 
%15" \citep{Giclas},
%13.4" in SDSS
%13.449" in 2Mass
$a_T= (898 \pm 188)$~AU, corresponding to an orbital period 
around ($28\,500 \pm 2200$)~years,
we get 
$\dot P_{\mathrm{orbital}} \leq (1.6 \pm 1.6) \times 10^{-15}\,{\mathrm{s/s}}$.
The large uncertainty takes into account the possibility the orbit might
be strongly elliptical.
\citet{Greaves}, in his section 4.3.1, discusses common proper motion pairs which possibly do not form a physical binary.
If G~117--B15A and B form a real binary system, the contribution of an orbital 
reflex motion to the observed $\dot P$
might account for one half of the observed $\dot{P}$.

\begin{figure}
\resizebox{\hsize}{!}{\includegraphics*[angle=0]{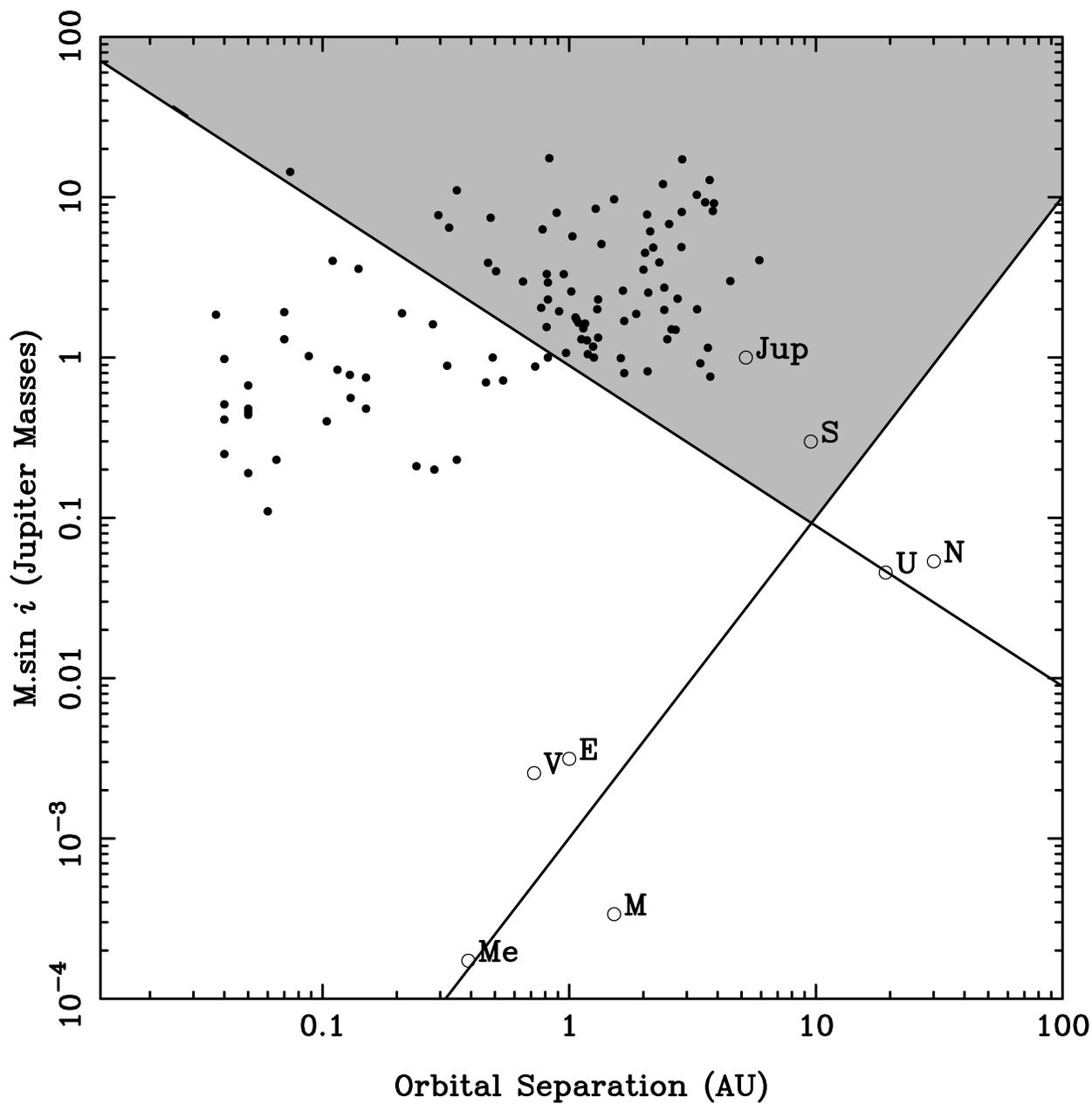}}
\caption{Planet exclusion region around G117-B15A.
The filled dots represent known extrasolar planets around
their stars, while the open circles represent solar system planets
around the Sun. The short period limit is the limit on the mass
of a planet that would produce a peak in the Fourier transform
smaller than those seen in Fig.~\ref{Figure2}. The long period
limit is the mass limit of a companion that would account
for the observed $\dot{P}$. Any planet above both lines would have
been detected.}
\label{fergal}
\end{figure}

The whole observed rate of period change 
{\it could} also be caused
by a planet of Jupiter's mass orbiting the white dwarf
at a distance of 31~AU, which
corresponds to an orbital period of 223~yr,
or a smaller planet on a closer orbit (see Figure~\ref{fergal}).
A planet with Jupiter's mass
any closer to the white dwarf would lead to a larger
$\dot{P}$.
\citet{Duncan} show that such a planet would survive the post-main
sequence mass loss.
Note that reflex motion produces periodic variations on
the $O-C$, which are distinguishable from parabolic variations
after a significant portion
of the orbit has been covered.

As discussed by \citet{Damour}, any relative acceleration of the
star with respect to the barycenter of the solar system will contribute
to the observed $\dot{P}$. Their equations (2.2) for the
differential galactic orbits, decomposed in
a planar contribution (2.12),
where the second term is the proper motion correction, 
and a perpendicular contribution (2.28),
applied to G~117-B15A, show the galactic contribution to be exactly the
one calculated above for proper motion, i.e., the other terms are
negligible (2 to 3 orders of magnitude smaller).

\section{Conclusions}
We have measured the rate of change of the main pulsation period
for the $T_{\mathrm{eff}} \simeq 12000$~K pulsating DA white dwarf
G~117--B15A, the first ZZ Ceti to have its evolutionary rate of
change measured, confirming it is the most stable optical clock
known, with a rate of change of 1~s in 8.9 million years and
a precise laboratory for physics at high energy.
It is important to notice that mode trapping,
a resonance between the local pulsation wavelength
with the thickness of one of the compositional layers, can reduce
the rate of period change by up to a factor of 2 \citep{Bradley96},
but the changes in the trapping layers are still caused
by cooling.

It has taken a huge investment of telescope time to achieve such
precision, but not only have we measured the cooling rate of this
400 million year old white dwarf \citep{Wood}, excluding the time the star
took to reach the white dwarf phase, we have also demonstrated
it does not harbor planetary bodies similar to Jupiter in mass
up to a distance around 30~AU from the star or smaller planets
with light travel time effects on the white dwarf larger than 1s.

We claim that the 215\,s periodicity in G~117-B15A is the most stable optical 
clock known.
\citet{Santra} and \citet{Hoyt} discuss projects to build
optical atomic clocks based on single trapped ions, or several
laser cooled neutron atoms, of strontrium or ytterbium,
expected to reach an accuracy of
$\dot{P}\leq 2 \times 10^{-17}$~s/s.
Considering their periods are $2.5\times 10^{-15}$~s and
$1.9 \times 10^{-13}$~s, even though they will be more accurate
than G~117-B15A, they will be much less stable, as their
timescales for period changes, $P/\dot{P}$, are 125~s and 3~h, compared
to 2~Gyr for G~117--B15A. Even the Hulse \& Taylor's millisecond pulsar
\citep{Hulse}, has a timescale for period change of only 0.35~Gyr
\citep{Damour}, but the radio millisecond pulsar
PSR~J1713+0747 \citep{Splaver} has $\dot{P}=8.1 \times 10^{-21}$~s/s,
and a timescale of 8~Gyr,
and PSR~B1885+09=PSR~J1857+0943 
with $\dot{P} = 1.78363 \times 10^{-20}$~s/s
has a stability timescale of 9.5 Gyr \citep{Kaspi}.
Together with these millisecond pulsars, G~117--B15A
may be used as the most stable 
known natural frequency and time-keeping standards
[e.g. \citet{KP}].
The white dwarf has the advantage of not having glitches and less
severe general relativistic corrections.

It is essential to note that the measured rate of period change
is consistent with the cooling rate in our white dwarf models
only for cores of C or C/O. With the refinement of the models,
the observed rate of period change can be used to accurately
measure the ratio of carbon to oxygen in the core of
this normal white dwarf, and therefore help 
in constraining the $C(\alpha,\gamma)O$ cross section.

%\section{Planet Formation}
%Alan Boss 
%[Alan Paul Boss (1951-), 2003, Astrophysical Journal, 599, 577]
%theory of planet formation, which predicts the planet
%forms at once from cloud fragmentation, must be correct 
%for the giant planets in the disk of spiral galaxies, because
%conservation of angular momentum requires a planet like
%Jupiter at Jupiter's distance to allow the formation of the
%central star like the Sun, remembering Jupiter accounts for 98\%
%of the solar system angular momentum.
%His models forms planets in around 1 Myr, the same timescale
%for the Sun to form.
%Inaba et al. (2003, Icarus, 166, 46) calculate the growth
%of the massive envelope onto a core takes at least $8\times 10^6$~yr.
%
%On the other hand, 
%[Shigeru Ida (1960-) \&
%Douglas N.C. Lin, 2004, Astrophysical Journal].
%theory of planetesimal aggregation
%must be correct for the smaller planets because they
%are well bellow the Jean's mass for fragmentation.

\acknowledgments
This work was partially supported by
grants from CNPq (Brazil), FINEP (Brazil), 
NSF (USA),  NASA (USA).
This research has made extensive use of
NASA's Astrophysical Data System Abstract Service.

%$$$$$$$$$$$$$$$$$$$$$$$$$$$$$$$$$$$$$$$$$$$$$$$$$$$
%                 TABLES.
%%%%%%%%%%%%%%%%%%%%%%%%%%%%%%%%%%%%%%%%%%%%%%%%%%%%

%\begin{table}
\begin{center}
\begin{deluxetable}{|r|l|c|c|}
\tablecaption{Total Data Set to Date}
\tablehead{
{\bf Time of Maximum} & {\bf Epoch of}   & {\bf (O-C)}  & {\bf $\sigma$}   \\
BJDD                 & {\bf Maximum}   &(sec)       & (sec) 
}
\startdata
2442397.917507 &       0 &  0.0 & 2.1 \\ 
2442477.797089 &   32071 &  0.5 & 1.7 \\ 
2442779.887934 &  153358 &  3.9 & 2.1 \\ 
2442783.850624 &  154949 &  1.2 & 2.9 \\ 
2442786.981458 &  156206 &  2.2 & 1.5 \\ 
2443462.962774 &  427607 &  1.6 & 1.4 \\ 
2443463.946592 &  428002 &  0.5 & 1.4 \\ 
2443465.969049 &  428814 &  0.5 & 1.6 \\ 
2443489.909755 &  438426 &  0.2 & 1.5 \\ 
2443492.898616 &  439626 &  0.9 & 1.6 \\ 
2443521.927837 &  451281 &  0.1 & 1.3 \\ 
2443552.752879 &  463657 &  0.8 & 1.4 \\ 
2443576.725940 &  473282 & -1.6 & 3.3 \\ 
2443581.692438 &  475276 &  0.3 & 1.3 \\ 
2443582.693698 &  475678 & -0.2 & 1.3 \\ 
2443583.697469 &  476081 &  1.0 & 1.3 \\ 
2443584.733602 &  476497 &  0.8 & 1.4 \\ 
2443604.659292 &  484497 &  1.3 & 1.5 \\ 
2443605.752703 &  484936 &  0.4 & 1.4 \\ 
2443611.693050 &  487321 &  0.6 & 1.3 \\ 
2443613.658222 &  488110 &  0.7 & 1.6 \\ 
2443636.674971 &  497351 &  8.8 & 3.4 \\ 
2443839.956765 &  578967 &  5.8 & 3.0 \\ 
2443841.976708 &  579778 &  3.7 & 3.5 \\ 
2443842.980413 &  580181 & -0.7 & 2.2 \\ 
2443843.944332 &  580568 &  0.5 & 2.6 \\ 
2443869.989703 &  591025 &  1.5 & 2.4 \\ 
2443870.946182 &  591409 &  5.5 & 3.1 \\ 
2443874.916339 &  593003 &  2.4 & 2.1 \\ 
2443959.695117 &  627041 &  0.1 & 2.0 \\ 
2443963.662836 &  628634 &  1.6 & 2.1 \\ 
2443990.664641 &  639475 &  2.7 & 1.3 \\ 
2444169.945954 &  711455 &  0.1 & 1.6 \\ 
2444231.822666 &  736298 & -0.7 & 2.9 \\ 
2444232.818992 &  736698 &  3.0 & 1.6 \\ 
2444293.833896 &  761195 &  0.3 & 1.8 \\ 
2444637.776174 &  899285 &  5.8 & 1.9 \\ 
2444641.624287 &  900830 &  2.8 & 1.1 \\ 
2444992.789531 & 1041820 &  0.1 & 1.6 \\ 
2444994.689956 & 1042583 &  1.2 & 1.2 \\ 
2444996.744801 & 1043408 &  2.0 & 1.3 \\ 
2444997.723649 & 1043801 &  1.9 & 1.2 \\
2445021.716661 & 1053434 &  1.7 & 1.4 \\ 
2445703.860004 & 1327309 &  1.9 & 1.7 \\ 
2445734.642701 & 1339668 &  2.4 & 1.2 \\ 
%\end{tabular}
%\end{center}
%\end{table}

%\begin{table}
%\begin{center}
%\begin{tabular}{|r|l|c|c|}  \hline
%{\bf Time of Maximum} & {\bf Epoch}   & {\bf (O-C)}  & {\bf $\sigma$}   \\
%BJDD                 &    &(sec)       & (sec) \\ \hline
2445735.643972 & 1340070 &  2.8 & 1.3 \\
2446113.763716 & 1491882 &  2.9 & 1.2 \\ 
2446443.775386 & 1624379 &  2.8 & 1.1 \\ 
2446468.630178 & 1634358 &  2.1 & 1.3 \\ 
2446473.718679 & 1636401 &  0.3 & 1.6 \\ 
2446523.620086 & 1656436 &  2.2 & 1.6 \\ 
2446524.613917 & 1656835 &  5.5 & 2.5 \\ 
2446768.855451 & 1754896 &  2.9 & 1.4 \\ 
2446794.935676 & 1765367 &  2.5 & 2.1 \\ 
2446796.928219 & 1766167 &  0.3 & 1.6 \\ 
2446797.924535 & 1766567 &  3.1 & 1.3 \\ 
2446798.903378 & 1766960 &  2.6 & 1.8 \\ 
2446823.663537 & 1776901 &  3.1 & 1.9 \\ 
2446825.651132 & 1777699 &  3.7 & 1.5 \\ 
2447231.328096 & 1940575 &  3.7 & 1.9 \\ 
2447231.612054 & 1940689 &  5.1 & 3.5 \\ 
2447232.396626 & 1941004 &  5.0 & 1.6 \\ 
2447232.623291 & 1941095 &  5.9 & 2.9 \\ 
2447233.343090 & 1941384 &  4.5 & 1.3 \\ 
2447233.634506 & 1941501 &  4.7 & 2.3 \\ 
2447234.319475 & 1941776 &  6.8 & 3.2 \\ 
2447235.313250 & 1942175 &  5.2 & 1.4 \\ 
2447235.607168 & 1942293 &  6.4 & 2.1 \\ 
2447236.610922 & 1942696 &  6.2 & 1.6 \\ 
2447589.375198 & 2084328 &  3.2 & 1.4 \\ 
2447594.331735 & 2086318 &  5.2 & 1.6 \\ 
2447595.323018 & 2086716 &  3.5 & 2.0 \\ 
2447596.311907 & 2087113 & 10.1 & 2.3 \\ 
2447597.315602 & 2087516 &  4.8 & 1.7 \\ 
2447598.319339 & 2087919 &  3.1 & 3.1 \\ 
2447499.072036 & 2048072 &  6.5 & 3.2 \\ 
2447532.768799 & 2061601 &  1.3 & 1.4 \\ 
2447853.846325 & 2190511 &  4.3 & 2.1 \\ 
2447856.832697 & 2191710 &  5.2 & 1.9 \\ 
2447918.644630 & 2216527 &  2.6 & 3.1 \\ 
2447920.619811 & 2217320 &  6.7 & 3.3 \\ 
2447952.622834 & 2230169 & -3.3 & 2.9 \\ 
2447972.620899 & 2238198 &  9.6 & 6.1 \\ 
2447973.709340 & 2238635 &  9.7 & 2.6 \\
2447973.741682 & 2238648 &  6.5 & 1.4 \\ 
2447978.770467 & 2240667 & 10.0 & 2.1 \\ 
2447979.781717 & 2241073 & 11.8 & 3.1 \\ 
2447980.319627 & 2241289 &  4.6 & 3.5 \\ 
2447977.403038 & 2240118 &  7.5 & 2.3 \\
2447978.327055 & 2240489 &  4.3 & 3.3 \\ 
2447979.358189 & 2240903 &  2.6 & 3.4 \\ 
2447979.358145 & 2240903 & -1.2 & 4.9 \\ 
2447978.601069 & 2240599 &  7.4 & 2.5 \\ 
%\end{tabular}
%\end{center}
%\end{table}

%\begin{table}
%\begin{center}
%\begin{tabular}{|r|l|c|c|}  \hline
%{\bf Time of Maximum} & {\bf Epoch}   & {\bf (O-C)}  & {\bf $\sigma$}   \\
%BJDD                 &    &(sec)       & (sec) \\ \hline
2447980.621017 & 2241410 &  5.8 & 3.4 \\ 
2447980.782929 & 2241475 &  7.2 & 2.3 \\ 
2447981.325918 & 2241693 &  8.4 & 1.4 \\ 
2447981.592393 & 2241800 &  5.7 & 1.4 \\ 
2447981.779185 & 2241875 &  4.8 & 1.1 \\ 
2447982.329663 & 2242096 &  7.4 & 1.8 \\ 
2447982.743093 & 2242262 &  5.0 & 1.2 \\ 
2447983.734400 & 2242660 &  5.4 & 1.2 \\ 
2447979.281057 & 2240872 &  9.5 & 2.9 \\ 
2447980.224899 & 2241251 & -2.4 & 2.9 \\ 
2447984.735678 & 2243062 &  6.5 & 1.1 \\ 
2448245.724666 & 2347847 & -3.3 & 5.1 \\ 
2448267.799932 & 2356710 &  5.2 & 2.3 \\ 
2448324.627972 & 2379526 &  4.3 & 1.2 \\ 
2448325.708938 & 2379960 &  4.1 & 1.3 \\ 
2448328.593208 & 2381118 &  6.4 & 1.6 \\ 
2448331.661735 & 2382350 &  4.0 & 1.2 \\ 
2448238.571479 & 2344975 &  8.3 & 2.2 \\ 
2448622.833258 & 2499253 &  3.3 & 1.8 \\ 
2448680.642683 & 2522463 &  6.3 & 1.2 \\ 
2448687.614155 & 2525262 &  4.0 & 1.2 \\ 
2448688.597979 & 2525657 &  3.4 & 1.2 \\ 
2449062.660365 & 2675840 &  4.2 & 1.6 \\
2449063.609354 & 2676221 &  6.7 & 1.9 \\ 
2449066.615640 & 2677428 &  6.5 & 1.4 \\ 
2449066.371558 & 2677330 &  7.2 & 2.0 \\ 
2449066.326737 & 2677312 &  8.2 & 2.6 \\ 
2449069.342967 & 2678523 &  6.4 & 1.7 \\ 
2449298.239287 & 2770423 &  8.5 & 4.1 \\ 
2449298.304041 & 2770449 &  8.2 & 4.1 \\ 
2449294.214264 & 2768807 &  5.5 & 4.1 \\ 
2449294.293897 & 2768839 & -0.5 & 4.1 \\ 
2449295.439583 & 2769299 & -4.0 & 6.1 \\ 
2449295.494387 & 2769321 & -3.3 & 7.1 \\ 
2449036.809260 & 2665461 &  2.4 & 2.2 \\ 
2449038.677300 & 2666211 &  3.1 & 2.2 \\ 
2449040.687310 & 2667018 &  3.6 & 4.1 \\ 
2449041.616360 & 2667391 &  4.9 & 4.1 \\ 
2449799.723888 & 2971765 &  5.6 & 1.3 \\ 
2450427.920960 & 3223981 &  8.2 & 3.8 \\ 
2450429.973242 & 3224805 &  2.7 & 2.4 \\ 
2450430.914779 & 3225183 &  6.9 & 2.5 \\ 
2450431.843821 & 3225556 &  7.5 & 1.5 \\ 
2450434.912392 & 3226788 &  8.8 & 2.0 \\ 
2450436.929828 & 3227598 &  5.4 & 1.7 \\ 
2450483.633189 & 3246349 &  9.6 & 1.8 \\
2451249.5989069 & 3553878&10.1&1.3 \\
2451249.7632895 & 3553944&9.7 &1.7 \\ 
%\end{tabular}
%\end{center}
%\end{table}

%\begin{table}
%\begin{center}
%\begin{tabular}{|r|l|c|c|}  \hline
%{\bf Time of Maximum} & {\bf Epoch}   & {\bf (O-C)}  & {\bf $\sigma$}   \\
%BJDD                 &    &(sec)       & (sec) \\ \hline
2451250.6126098 & 3554285&8.7&2.0 \\
2451526.8772586 &3665203& 10.4& 1.2\\
2451528.8523866 &3665996&10.0&1.5\\
2451528.9196061& 3666023&7.4&1.4\\
2451528.9868422& 3666050&6.3&1.9\\
2451529.8585943& 3666400& 6.6&2.0\\
2451530.9097492 & 3666822&    13.4&2.41\\
2451960.8561629 & 3839442&    10.1&1.62\\
2451962.7864775 & 3840217&    11.3&1.48\\
2451967.6806926 & 3842182&    8.1 &1.93\\
2451988.7919772 & 3850658&   10.5 &2.00\\
2451990.7845255 & 3851458&    8.8 &1.59\\
2452037.6472583 & 3870273&   10.1 &3.39\\
2452045.6399770 & 3873482&    12.5&1.85\\
2452225.9050927 & 3945857&    7.6 &1.34\\
2452225.9598927 & 3945879&    8.0 &0.65\\
2452263.8834810 & 3961105&    10.6&0.58\\
2452316.6442205 & 3980721&    13.1&1.0\\
2452317.8995164 & 3982288&    12.2&0.67\\
2452319.7999417 & 3982691&    12.0&0.79\\
2452317.6479750 & 3982792&    10.3&0.95\\
2452321.8348344 & 3983555&    11.4&1.23\\
2452322.7265266 & 3984372&    9.9 &3.03\\
2452312.7412881 & 3984730&    11.4&3.5\\
2452373.6840808 & 4089983&     9.9&1.1\\
2452373.6839702 & 4090425&    12.6&1.2\\
2452373.7140655 & 4090791&    10.6&0.68\\
2452375.6392709 & 4122465&    12.8&1.0\\
2452374.7700070 & 4124076&    12.5&1.23\\
2452581.9494464 & 4134940&  12.0  &1.83\\
2452583.9095168 & 4137288&  11.5  &1.74\\
2452584.8812628 & 4144503&  13.8  &0.92\\
2452585.9821875 & 4148953&  12.0  &1.16\\
2452586.8937641 & 4146104&  12.3  &0.7\\
2452665.7845581 & 4255806&  13.7  &0.72\\
2452669.7970851 & 4256260&  10.6  &1.52\\
2452696.8561592 & 4257768&  12.8  &0.96\\
2452724.6624548 & 4267394&  10.5  &0.81\\
2453381.7442572 & 4409917&  11.5  &1.41\\
2453439.7653860 & 4433212&  13.9  &1.0\\
2453446.6073830 & 4435959&  15.2  &0.71\\
2453473.6191370 & 4446804&  15.1  &0.75\\
\enddata
\end{deluxetable}
\label{tmax}
\end{center}
%\end{table}

\end{document}